\documentclass[10pt,aps,twocolumn,floats,floatfix,amssymb,prl,superscriptaddress]{revtex4-2}

\usepackage{graphicx}
\usepackage{amsmath,amssymb}
\usepackage{amsfonts}
\usepackage{xspace} 
\usepackage{xcolor}
\usepackage{dcolumn}
\usepackage{bm}
\usepackage{mathrsfs}
\definecolor{Burgundy}{RGB}{144,0,32}
\usepackage[colorlinks=true,linkcolor=Burgundy,citecolor=Burgundy, filecolor=Burgundy, urlcolor=Burgundy]{hyperref}
\usepackage[all]{hypcap} 
\usepackage[utf8]{inputenc} 
\usepackage{multirow}
\usepackage{etoolbox}
\usepackage{tikz}

\newcommand{\dd}{\mathrm{d}}

\begin{document}

\title{
Regular black holes without mass-inflation instability and gravastars\\from modified gravity
}

\author{Astrid Eichhorn}
\email{eichhorn@thphys.uni-heidelberg.de}
\affiliation{Institut f\"ur Theoretische Physik, Universit\"at Heidelberg, Philosophenweg 12 \& 16, 69120 Heidelberg, Germany}

\author{Pedro G.~S.~Fernandes}
\email{fernandes@thphys.uni-heidelberg.de}
\affiliation{Institut f\"ur Theoretische Physik, Universit\"at Heidelberg, Philosophenweg 12 \& 16, 69120 Heidelberg, Germany}

\begin{abstract}
We derive regular black-hole solutions, including the Hayward metric, from four-dimensional action principles involving vector fields in addition to the metric. These black holes possess additional hair associated with the vector fields, manifesting as free integration constants that regularize the geometry. These constants can be chosen such that regular black holes of all masses are extremal. As a result, they have vanishing surface gravity and are \emph{not} susceptible to mass-inflation instability. We also discover another regular black-hole metric with these properties, which constitutes a gravastar for an appropriate choice of integration constant.
\end{abstract}

\maketitle

\noindent \textbf{\textit{Introduction.}} Singularities are a generic feature of solutions in General Relativity (GR) \cite{Hawking:1970zqf,Penrose:1964wq,Hawking:1967ju}. 
At a curvature singularity, physical quantities such as tidal forces diverge, signalling a breakdown of the theory. As a consequence, the Kerr metric, despite fitting experimental data from gravitational-wave observatories \cite{LIGOScientific:2016aoc,LIGOScientific:2016lio,LIGOScientific:2019fpa,LIGOScientific:2020tif,LIGOScientific:2021sio} and very-large-baseline-interferometry \cite{EventHorizonTelescope:2019dse,EventHorizonTelescope:2022wkp,EventHorizonTelescope:2022xqj}, cannot be more than an approximate description of the exterior spacetime geometry of a black hole (BH).

\begin{figure}[!t]
	\centering
	\includegraphics[width=0.9\linewidth,clip=true, trim=0cm 0.5cm 0.5cm 1cm]{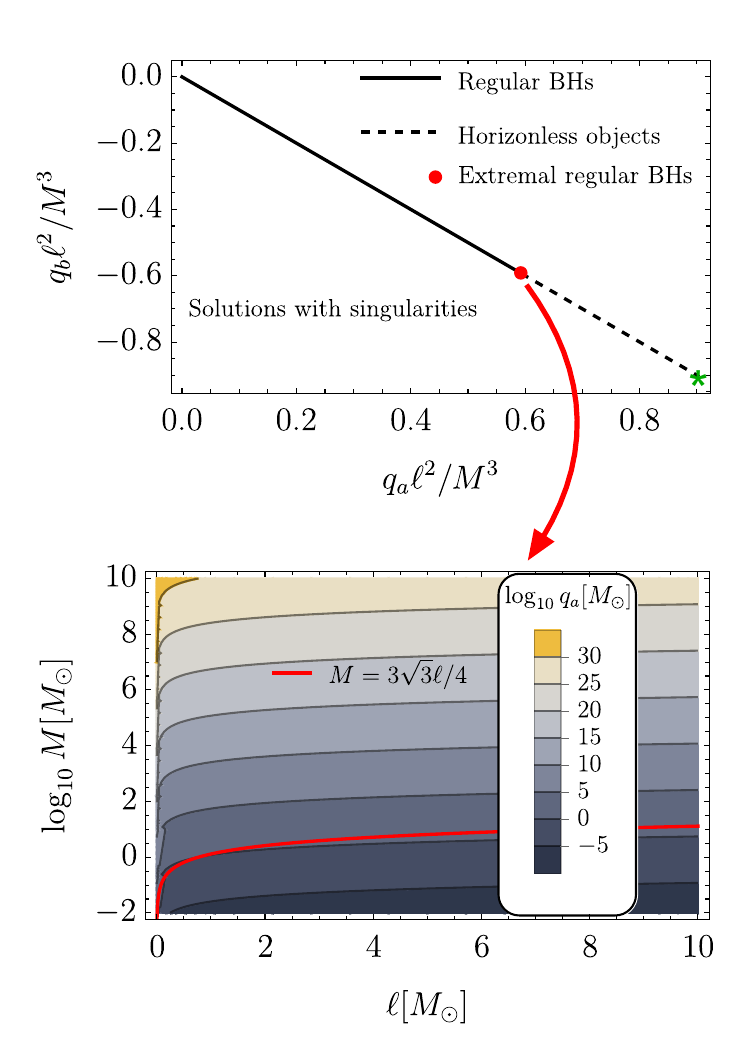}\hfill
    \caption{
    Upper panel: Our solutions depend on two integration constants, $q_a$ and $q_b$, here combined with the BH mass $M$ and the beyond-GR scale $\ell$ into dimensionless quantities. For $q_b = -q_a$, the BHs are regular, and for $q_a = 16M^3/(27\ell^2)$ they are extremal. The green star marks the largest $|q_{a,b}|$ for which horizonless objects have light rings. Lower panel: The Hayward extremality bound $M = (3 \sqrt{3} \ell)/4$ (red line) sets a limit below which no BHs exist described by Eq.~\eqref{eq:hayward}.
    In contrast, in our theory, choosing $q_a$ as indicated by the contours, yields an extremal, regular BH for any $M$ and $\ell$.
    }
    \label{fig:parameterspace}
\end{figure}

BH spacetimes must be regular, if differential geometry and field theory are adequate tools to describe the BH interior. 
For a regular BH, physical quantities are finite, see, e.g., \cite{1968qtr..conf...87B,Hayward:2005gi,Dymnikova:1992ux,Bonanno:2000ep,Modesto:2008im,Platania:2019kyx,Bambi:2013ufa,Toshmatov:2014nya,Azreg-Ainou:2014pra,Simpson:2019mud,Eichhorn:2021etc,Lan:2023cvz,Carballo-Rubio:2025fnc,Torres:2022twv,Carballo-Rubio:2023mvr}. A minimal example of such a BH is the Hayward metric \cite{Hayward:2005gi}, defined by the line element 
\begin{eqnarray}
    \dd s^2 &=& -f(r) \dd v^2 + 2\dd v \dd r + r^2\left( \dd \theta^2 + \sin^2\theta \dd \varphi^2 \right),
    \label{eq:sphericalsymmetry}\\
&{}&\mbox{with }\quad
    f(r) = 1 - \frac{2M r^2}{r^3 + 2M \ell^2}.
    \label{eq:hayward}
\end{eqnarray}
The length scale $\ell$ is associated with beyond-GR effects. It limits the spacetime curvature so that, e.g., the square of the Weyl tensor becomes $C_{\mu\nu\rho\sigma}C^{\mu\nu\rho\sigma} = 48 M^2 r^6(r^3-4M\ell^2)^2/(2M \ell^2 +r^3)^6$ and is no longer divergent at $r\rightarrow 0$.

Our understanding of regular BHs faces several challenges, see \cite{Carballo-Rubio:2025fnc}. Chief among them is that regular BHs are rarely found as solutions to the field equations of a well-defined gravitational theory. For instance, BH solutions of GR coupled to non-linear electrodynamics are typically only regular for a special value of the ratio of the BH mass to the coupling constant. Thus, the BH mass appears in the Lagrangian
\cite{Ayon-Beato:1998hmi,Bronnikov:2000vy,Ayon-Beato:2000mjt,Dymnikova:2004zc,Bronnikov:2017sgg,Huang:2025uhv}. In addition, these solutions are not stable \cite{DeFelice:2024seu}.
Finding a theory in 3+1 spacetime dimensions, in which BHs can be regular for any value of their mass has long been an open challenge. Despite recent attempts \cite{Cano:2020qhy, Cano:2020ezi, Babichev:2020qpr, Baake:2021jzv, Colleaux:2017ibe, buenoRegularBlackHoles2021, buenoRegularChargedBlack2025,Giacchini:2024exc, Bueno:2024dgm,Fernandes:2025eoc,Fernandes:2025fnz}, e.g., based on infinite towers of beyond-GR terms \cite{Bueno:2024dgm,Fernandes:2025eoc,Fernandes:2025fnz,Bueno:2024zsx,Bueno:2024eig,Bueno:2025gjg,Hennigar:2025ftm,Aguayo:2025xfi,Frolov:2024hhe,DiFilippo:2024mwm,Konoplya:2024hfg,Cisterna:2025vxk}, as well as phenomenological models from quantum gravity \cite{Bonanno:2000ep,Eichhorn:2022bgu,Platania:2023srt,Ashtekar:2023cod,Giacchini:2023waa}, this challenge remains open.

A second challenge concerns the stability of regular BHs. Generically, they have an inner horizon \cite{Carballo-Rubio:2019nel,Carballo-Rubio:2019fnb} with non-zero surface gravity. Close to the inner horizon, gravitational energy grows exponentially, a phenomenon known as mass inflation \cite{PhysRevLett.63.1663,PhysRevLett.67.789,Hamilton:2008zz,Brown:2011tv,Carballo-Rubio:2018pmi,Bertipagani:2020awe,Bonanno:2020fgp,Carballo-Rubio:2021bpr,Hale:2025ezt}.
This indicates an instability, though its growth may be limited \cite{Bonanno:2022rvo,Bonanno:2025bgc}. The instability can be entirely avoided if the inner horizon is extremal \cite{Carballo-Rubio:2022kad,Franzin:2022wai} (see however \cite{McMaken:2023uue}), or if BHs are fully extremal \cite{DiFilippo:2024spj}.
This can be achieved in regular BHs for a fixed ratio of $\ell$ to $M$, e.g., for the Hayward metric, only BHs with $M=(3 \sqrt{3} \ell)/4$ are extremal.
It is an open challenge to find a theory in which non-spinning BHs of any value of the mass can be extremal.

We propose a framework in which both challenges can be met: we introduce vector-tensor theories, whose spherically symmetric solutions are characterized by primary hair \cite{Charmousis:2025jpx,Bakopoulos:2023fmv,Khodadi:2020jij,Vagnozzi:2022moj,Herdeiro:2014goa,Herdeiro:2016tmi,Bakopoulos:2023fmv,Baake:2023zsq,Bakopoulos:2023sdm,Charmousis:2025xug,Myung:2025afs} which can be adjusted to render BH solutions regular and extremal across all masses, cf.~Fig.~\ref{fig:parameterspace}.\\

\noindent \textbf{\textit{The theory.}}
We consider the action\footnote{We work in units in which $G=1=c$.} 
\begin{equation}
    S = \frac{1}{16\pi} \int \dd^4 x \sqrt{-g} \left( R + \ell^2 (\mathcal{L}[A]-\mathcal{L}[B]) \right),
    \label{eq:action}
\end{equation}
where we have defined the vector-tensor Lagrangian as
\begin{equation}
    \mathcal{L}[W] =  4G^{\mu \nu}W_\mu W_\nu + 8 W_\nu W^\nu \nabla_\mu W^\mu + 6(W_\nu W^\nu)^2,
    \label{eq:VTtheory}
\end{equation}
This theory is a generalized (bi-)Proca theory \cite{Heisenberg:2014rta}, and we present its field equations in the \hyperref[suppmaterial]{Supplemental Material}.
Its roots lie in a dimensional regularization of the Gauss-Bonnet invariant, $\mathcal{G} = R^2 - 4 R_{\mu \nu} R^{\mu \nu} + R_{\mu \nu \alpha \beta} R^{\mu \nu \alpha \beta}$, which is a higher-curvature term expected from proposed UV completions of gravity, such as string theory \cite{Gross:1986mw,Gross:1986iv,Grisaru:1986vi} and asymptotically safe quantum gravity \cite{Codello:2006in,Ohta:2013uca,Ohta:2015zwa,Knorr:2018kog,Falls:2020qhj,Knorr:2022dsx}, and is motivated as well by fundamental results in gravitational theory, such as Lovelock's theorem~\cite{lovelockEinsteinTensorIts1971}. As a topological invariant in $d=4$ dimensions, $\mathcal{G}$ does not contribute to the equations of motion. In \cite{Glavan:2019inb}, it was proposed that this changes for a coupling scaling as $\sim (d-4)^{-1}$, and in \cite{Fernandes:2020nbq,Lu:2020iav,Kobayashi:2020wqy,Hennigar:2020lsl}, this idea was first implemented consistently, resulting in scalar-tensor theories, see \cite{Fernandes:2022zrq} for a review.

The regularization of the Gauss-Bonnet term to four dimensions can also be done starting from a Gauss-Bonnet-scalar constructed from a Weyl connection \cite{Charmousis:2025jpx}, with an associated covariant derivative $\tilde{\nabla}_\mu$, such that $\tilde{\nabla}_{\lambda}g_{\mu \nu}=-2g_{\mu\nu}W_{\lambda}$. The resulting Weyl vector $W_{\mu}$ persists in the limit $\lim_{d\to4} (\mathcal{G} - \tilde{\mathcal{G}}[W])/(d-4)$, where $\tilde{\mathcal{G}}[W]$ is the Gauss-Bonnet scalar constructed from the Weyl connection which is expressed as
\begin{equation}
    \tilde{\mathcal{G}}[W]= \mathcal{G} + (d-3)\nabla_\mu J^\mu[W] + (d-3)(d-4) \mathcal{L}[W],
    \label{eq:GB-Weyl}
\end{equation}
where
\begin{equation}
    \begin{aligned}
            &J^\mu[W] =  8 G^{\mu \nu} W_\nu + 4(d-2)\left(W^2 + \nabla_\nu W^\nu - W_\nu \nabla^\nu\right)W^\mu,\\&
    \mathcal{L}[W] = 4 G^{\mu \nu}W_\mu W_\nu + (d-2)\left( 4 W^2 \nabla_\mu W^\mu + (d-1)W^4 \right),
    \end{aligned}
\end{equation}
see Refs.~\cite{Charmousis:2025jpx,jimenezExtendedGaussBonnetGravities2014, bahamondeExactFiveDimensional2025} for a detailed derivation of the expression for the Gauss-Bonnet-scalar. The vector field provides primary hair: a new constant of integration in BH geometries~\cite{Charmousis:2025jpx}.

Here, we generalize this theory further, and obtain a new regularized four-dimensional Gauss-Bonnet theory, by introducing two Weyl vectors $A_{\mu}$ and $B_{\mu}$. These should be considered as auxiliary fields, because we do not couple matter to the corresponding connection. Using Eq. \eqref{eq:GB-Weyl}, we obtain the vector-tensor part of the action~\eqref{eq:action} in the limit
\begin{equation}
    \mathcal{L}[A]-\mathcal{L}[B] = \lim_{d\to 4} \frac{\tilde{\mathcal{G}}[A]-\tilde{\mathcal{G}}[B]}{d-4}.
    \label{eq:regularization}
\end{equation}
Neither of the two vector fields has a kinetic term; they should therefore be thought of as (non-linear) constraints. In fact, for the BH solution we consider below, it holds that both vector fields are null; thus, the solution is insensitive to the term $(W_{\nu}W^{\nu})^2$ in $\mathcal{L}[W]$.\\

\noindent \textbf{\textit{The Hayward metric as an exact solution.}}
To derive static, spherically symmetric BH solutions to the theory \eqref{eq:action}, we consider a line element in ingoing null coordinates given by Eq.~\eqref{eq:sphericalsymmetry} and restrict the vector fields to have spherically symmetric, time-independent profiles, $A_\mu \dd x^\mu = a(r) \dd v$, and $B_\mu \dd x^\mu = b(r) \dd v$. 
Following the method of Ref.~\cite{Charmousis:2025jpx}, we find that
\begin{eqnarray}
    a(r) &=& \frac{r - r f(r) - q_a/2}{2r^2},\,\, b(r) = \frac{r - r f(r) -q_b/2}{2r^2},
    \label{eq:VFprofiles}\\
    f(r) &=& 1 - \frac{8M r^3 + \ell^2 (q_a^2-q_b^2)}{4r(r^3+ (q_a-q_b) \ell^2)},
    \label{eq:metric_sing}
\end{eqnarray}
solve the equations of motion \footnote{We provide a Mathematica notebook with the field equations and a detailed derivation of the solution as an ancillary file.}. Here, $M$, $q_a$ and $q_b$ are  constants of integration. $M$ is the ADM-mass of the BH, while the other constants are primary hairs: quantities that modify the geometry and are not fixed in terms of the BH mass.

Near $r=0$, the metric function \eqref{eq:metric_sing} behaves as
\begin{equation}
    f(r) = -\frac{q_a + q_b}{4r} + 1 - \frac{8M -q_a - q_b}{4\ell^2 (q_a-q_b)} r^2 + \mathcal{O}(r^3).
\end{equation}
For regular BHs, for which non-derivative curvature invariants are everywhere finite,
$f(r)$ must be of the form $f(r) = 1 + \mathcal{O}(r^2)$. To achieve this, we must set $q_b = -q_a$, such that the metric function becomes
\begin{equation}
    f(r) = 1 - \frac{2M r^2}{r^3+2 q_a \ell^2}.
    \label{eq:regular_bh}
\end{equation}
This is a Hayward-type metric \cite{Hayward:2005gi}, where the singularity is regularized by a free primary hair parameter $q_a$. If the integration constant is fixed such that $q_a=M$, the original Hayward metric in Eq.~\eqref{eq:hayward} is obtained. However, this is only one point in the parameter space of regular BH solutions, see the upper panel in Fig.~\ref{fig:parameterspace}.
We take 
$q_a>0$ to avoid singularities at positive $r$.
Radial null geodesics can be continued across $r=0$, if we make the physically acceptable choice that the sixth derivative of the geodesic is not continuous at $r=0$ \cite{Zhou:2022yio,Torres:2022twv}.

The mechanism behind regularity is a violation of the strong energy condition by the combination of $A_{\mu}$ and $B_{\mu}$, which occurs for $r< (q_a \ell^2)^{1/3}$, being confined to a small region near the core, entirely within the event horizon. Since no divergences in physical quantities occur, the violation of the strong energy condition is controlled. 
In fact, the components of the stress-energy tensor corresponding to each vector field 
diverge at the origin (see the \hyperref[suppmaterial]{Supplemental Material}). The divergent contributions have opposite signs for $q_b= -q_a$ and hence cancel, leaving behind a finite part that violates the strong energy condition, and sources a regular core. This is consistent with the well-known result that a regular BH core necessarily violates the strong energy condition \cite{Zaslavskii:2010qz}.

The fixing of the integration constants $q_a$ and $q_b$ should not be considered fine-tuning. Similar to situations in other areas of physics, imposing physical conditions (e.g., regularity), on the space of mathematical solutions fixes free parameters to particular values. For instance, in scalar-Gauss-Bonnet gravity, the scalar charge is fixed in terms of the BH mass in order to have a regular solution on the horizon \cite{Sotiriou:2013qea,Sotiriou:2014pfa}. 

At a phenomenological level, the ``decoupling" of the hair $q_a$ from the fundamental length scale $\ell$ of the theory has important consequences. In the enlarged BH parameter space spanned by $M$ and $q_a$, and at fixed $\ell$, BHs of \emph{any} mass can exhibit large deviations from Schwarzschild BHs, if the value of $q_a$ for a given BH is chosen large enough. This is distinct from the usual Hayward case, where deviations from Schwarzschild decrease monotonically with increasing BH mass $M$, for fixed $\ell$. Observational tests of BH geometries that in combination cover a large range of BH masses, such as, e.g., tests with the Ligo-Virgo-Kagra-interferometers \cite{LIGOScientific:2016aoc,LIGOScientific:2016lio,LIGOScientific:2019fpa,LIGOScientific:2020tif,LIGOScientific:2021sio} and the Event Horizon Telescope \cite{EventHorizonTelescope:2019dse,EventHorizonTelescope:2021dqv,EventHorizonTelescope:2022wkp,EventHorizonTelescope:2022xqj}, are paramount to test this and similar theories.
 
At this point, we have addressed the first challenge of regular BHs, namely obtaining a regular solution at all ratios of BH mass to the coupling constant.\\

\noindent\textbf{\textit{Achieving extremality at all ratios of mass to coupling.}}
In the regime where the regular geometry \eqref{eq:regular_bh} describes a BH, the inner horizon is a Cauchy horizon \cite{Chiba:2017nml}. When the Cauchy horizon has a non-zero surface gravity, we expect the BH to suffer from mass-inflation instability. 
We avoid the mass-inflation instability by fixing
\begin{equation}
q_a = \frac{16}{27} \frac{M^3}{\ell^2}.\label{eq:qacrit}
\end{equation}
This renders the geometry extremal, so that it has zero surface gravity and contains a single (degenerate) event horizon located at $r = 4M/3$ and the metric function \eqref{eq:regular_bh} becomes 
\begin{equation}
    f(r) = 1 - \frac{2M r^2}{r^3 + \frac{32}{27}M^3}.
    \label{eq:regular_extremal}
\end{equation}
This describes a regular extremal BH for all values of the BH mass $M$. This scenario corresponds to the red point in the upper panel in Fig.~\ref{fig:parameterspace}.  The coupling constant $\ell$ of the theory no longer appears in the metric function. However, unlike in the case of extremal Hayward, this is \emph{not} because of a relation between $\ell$ and $M$, cf.~lower panel of Fig.~\ref{fig:parameterspace}.

Thereby, we have now addressed the second challenge of regular BHs: finding regular BH solutions free from the mass inflation instability for all values of the BH mass.\\

\noindent \textbf{\textit{Horizonless spacetimes.}}
Linked to the above challenges for BHs, there is a question whether astrophysical observations can all be explained by BH mimickers, which are (ultra-) compact objects without a horizon \cite{Broderick:2005xa,Carballo-Rubio:2022imz}, typically assumed to be regular, see \cite{Cardoso:2019rvt} for a review. Similarly to the case of extremal regular BHs, such horizonless spacetimes  typically require us to choose the BH mass smaller than a critical value, set by the coupling constant of the theory. For instance, for the Hayward spacetime, $M< 3\sqrt{3}\ell/4$, is required to render the spacetime horizonless \cite{Carballo-Rubio:2022nuj}. The situation is again different in our theory, where $q_a>\frac{16}{27} \frac{M^3}{\ell^2}$ results in a horizonless spacetime, which is achievable for all combinations of mass and coupling constant. 

When compact enough ($q_a < 3125M^3/(3456\ell^2)$, see upper panel of Fig.~\ref{fig:parameterspace}), these horizonless spacetimes have two light rings, the inner of which is stable.
Because of this, these horizonless compact objects might be subject to a light-ring instability \cite{Cunha:2017qtt,Cunha:2022gde,Cardoso:2014sna,Keir:2014oka,Cunha:2025oeu,Siemonsen:2024snb,Marks:2025jpt,Benomio:2024lev,Redondo-Yuste:2025hlv,Guo:2024cts,Franzin:2023slm}, caused by the accumulation of massless perturbations and their eventual backreaction at the light ring. This instability is non-linear and model-dependent.\\

\noindent \textbf{\textit{Another regular black hole geometry as a solution.}}
We consider  a generalization of the action \eqref{eq:action}
with a relative coupling between the vector-tensor Lagrangians: $(\mathcal{L}[A]-\mathcal{L}[B]) \to (\mathcal{L}[A] - \kappa \mathcal{L}[B])$, such that the theory in Eq.~\eqref{eq:action} is recovered for $\kappa=1$.
When $\kappa \neq 1$, a larger family of regular BH solutions exists, which, in addition to $q_a$, is characterized by the coupling $\kappa$. 

Choosing the same profiles as before for the vector fields, but fixing $q_b=-q_a/\sqrt{\kappa}$, we obtain the following asymptotically-flat regular BH solution\footnote{As in standard Gauss-Bonnet theories, flipping the sign in front of the square root yields a non-asymptotically flat branch, which is also a solution to the equations of motion. When $Q=0$, the standard (non-regular) geometry of 4D-Gauss-Bonnet gravity BHs is recovered \cite{Fernandes:2022zrq}.}
\begin{equation}
\small
    f(r) = 1 - \frac{M Q}{r \lambda} - \frac{r^2}{2 \lambda  M^2} \left(1-\sqrt{\left(1+\frac{2 M^3 Q}{r^3}\right)^2-\frac{8 \lambda  M^3}{r^3}}\right),
    \label{eq:newregular}
\end{equation}
where we have defined the dimensionless quantities $Q \equiv (q_a \ell^2 (1+\sqrt{\kappa}))/(2M^3)$, and $\lambda \equiv (1-\kappa) \ell^2/M^2$.
An analysis of the Kretschmann scalar (see the \hyperref[suppmaterial]{Supplemental Material}) reveals that the spacetime is everywhere regular as long as $Q>\max\{\lambda,0\}$\footnote{We use that for static, spherically symmetric solutions, the full basis of non-derivative curvature invariants \cite{Zakhary:1997xas} remains finite for $r\geq 0$, if the Kretschmann scalar remains finite, because it is a sum of squares of the independent components of the Riemann tensor \cite{Simpson:2023apa,dePaulaNetto:2023cjw}}.

\begin{figure}[!t]
	\centering
	\includegraphics[width=\linewidth]{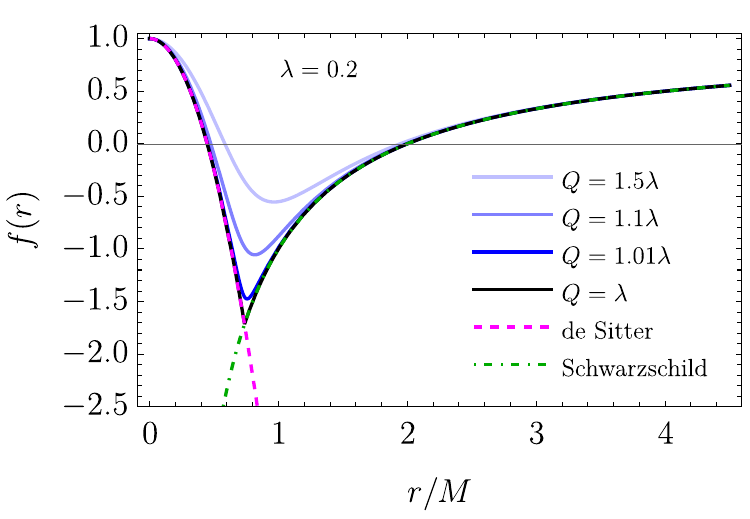}\hfill
    \includegraphics[width=\linewidth]{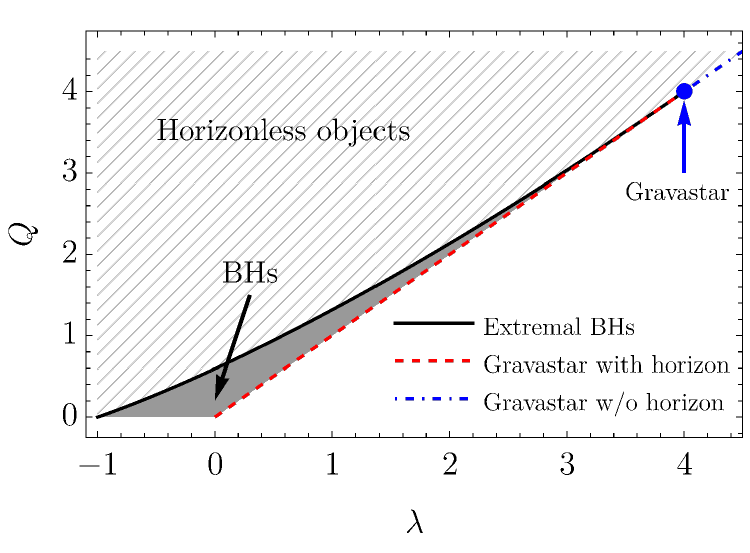}\vfill
    \caption{Upper panel: The metric function in Eq.~\eqref{eq:newregular} describes a regular object that reduces to a gravastar in the limit $Q = \lambda$, and smoothly departs from this regime as $Q > \lambda$ increases. Lower panel: space of solutions in the $\lambda-Q$ plane.}
    \label{fig:gravastar}
\end{figure}

The metric function \eqref{eq:newregular} describes a regular BH with event and inner horizons located at
\begin{equation}
\small
    \begin{aligned}
        &r_+ = \frac{2M}{3}\left[ 1 + \sqrt{4+3\lambda} \cos \left( \frac{1}{3} \cos^{-1}\left( \frac{8-27Q+9\lambda}{\left(4+3\lambda\right)^{3/2}} \right) \right)\right],\\&
        r_-= \frac{2M}{3}\left[ 1 - \sqrt{4+3\lambda} \sin \left( \frac{1}{3} \sin^{-1}\left( \frac{8-27Q+9\lambda}{\left(4+3\lambda\right)^{3/2}} \right) \right)\right],
    \end{aligned}
    \label{eq:horizons}
\end{equation}
respectively, if the following conditions hold:
\begin{equation}
    \max\{\lambda,0\} <Q \leq Q_{\rm crit} \equiv \frac{1}{27} \left( 8+9\lambda +  \left(4+3\lambda\right)^{3/2} \right),
    \label{eq:alphaBH}
\end{equation}
and $-1< \lambda < 4$. 
In the case $Q=Q_{\rm crit}$, the regular BH is extremal. Given a set of coupling constants $\ell, \kappa$, as well as a BH mass, the extremality condition fixes $q_a$.\\

\noindent \textbf{\textit{Gravastar as a solution.}}
Besides the choice $Q = Q_{\rm crit}$, a second choice of $Q$ is phenomenologically interesting, namely $Q \to \lambda$, when $\lambda>0$.
The function $f(r)$ has the asymptotic expansions $f(r) \sim 1- \frac{r^2}{Q\, M^2}+\mathcal{O}(r^3)$, corresponding to a de Sitter core, and $f(r) \sim 1- \frac{2M}{r}+ \mathcal{O}(r^{-2})$, corresponding to a Schwarzschild exterior. 
The interpolation between the two limits is determined by the constant of integration, $Q$, and the coupling constants that determine $\lambda$. As $Q \rightarrow  \lambda$, the transition between the two regimes becomes increasingly sharp, see upper panel of Fig.~\ref{fig:gravastar}.
Therefore, our theory provides a concrete realization of an object similar to a gravastar \cite{Mazur:2001fv, Visser:2003ge}, without the need to invoke a thin shell \cite{Cattoen:2005he,DeBenedictis:2005vp} of unknown matter.
In the limit $Q=\lambda$, the metric function in Eq.~\eqref{eq:newregular} joins the Schwarzschild exterior to the de Sitter core directly, without an interpolating region,
\begin{equation}
    f(r)\rvert_{Q=\lambda} = \begin{cases}
    1-\frac{2M}{r}, & \text{if } r \geq M(2\lambda)^{1/3}\\
    1 - \frac{1}{\lambda\, M^2} r^2 & \text{otherwise}.
\end{cases}
\label{eq:gravastar}
\end{equation}
The radius at which the transition occurs is determined by $\lambda$ and is given by $r_{\rm trans}=M (2\lambda)^{1/3}$, which lies inside the event horizon for $\lambda<4$ (cf.~Fig.~\ref{fig:gravastar}). 
For $\lambda=4$, however, the transition point coincides with the Schwarzschild radius, much like for the original gravastar proposals \cite{Mazur:2001fv, Visser:2003ge}. For a fixed value of the coupling constant of the theory, such a gravastar can be realized for a specific value of the mass, whereas other masses constitute a generalization of the original gravastar proposals.
The space of solutions in the $\lambda-Q$ plane is shown in the lower panel of Fig.~\ref{fig:gravastar}.

To the best of our knowledge, this is the first time the regular geometry presented in Eq.~\eqref{eq:newregular} is discussed, and also the first time a gravastar-like object is obtained as a solution to the dynamics of a theory, in which the field content is fully specified \footnote{Solutions with properties similar to gravastars were derived in \cite{Carballo-Rubio:2017tlh} in the context of stellar equilibrium in semiclassical gravity.}.
\\

\noindent \textbf{\textit{Discussion.}}
In this work, we have presented a four-dimensional theory in which the spherically symmetric BH solutions feature two constants of integration. One of these can be used to avoid the spacetime singularity at $r=0$ and render all non-derivative curvature invariants finite.
The second constant of integration can be used to avoid the inner-horizon instability by making the BH extremal. Importantly, this can be achieved for \emph{any} combination of BH mass and coupling constants of the theory.

Our work offers a potential realization of the proposal put forward in Ref.~\cite{DiFilippo:2024spj}, which argues that, when accounting for known spacetime instabilities, such as mass inflation and light ring instabilities, regular extremal BHs should be the stable endpoint of gravitational collapse.

A potential issue in this construction is that extremal BHs might be subject to the Aretakis instability \cite{Aretakis:2011ha,Aretakis:2011hc,Aretakis:2012ei,Murata:2013daa,Lucietti:2012xr}. This is a generic instability of extremal horizons under perturbations, and in the case of an extremal Reissner-Nordstr\"om BH, it was shown that the endpoint of the instability is a non-extremal BH \cite{Murata:2013daa}. Understanding the behavior of the solution under perturbations is an important avenue for future research. Similarly, understanding the sensitivity of the extremal solution to higher-order curvature corrections is pertinent \cite{Horowitz:2023xyl,Horowitz:2024dch}.

Since in our scenario BHs are extremal, they are dark-matter candidates, as no BH can evaporate via Hawking radiation \cite{Hawking:1975vcx}, regardless of its mass. Small enough extremal BHs would act as cold relics. Scenarios where BH remnants are proposed as dark matter have been put forward, e.g., in Refs.~\cite{macgibbonCanPlanckmassRelics1987,DiGennaro:2021vev,Rasanen:2018fom,Fernandes:2021ysi,PhysRevD.46.645,Green:1997sz,Alexeyev:2002tg,Chen:2004ft,Nozari:2005ah,Carr:1994ar,Lehmann:2019zgt,Bai:2019zcd,Lehmann:2021ijf,Calza:2024fzo,Calza:2024xdh,Santiago:2025rzb,Davies:2024ysj}, see also Refs.~\cite{Chen:2014jwq,Ong:2024dnr} for reviews. In our case, due to the lack of Hawking radiation, exclusion limits on low-mass black holes must be re-evaluated.

Moreover, since these BHs do not evaporate, there is no loss of information in Hawking radiation, and thus no information paradox \cite{Hawking:1975vcx,Hawking:1976ra,Mathur:2009hf,Almheiri:2012rt}. Similar considerations apply to the no-global-symmetries conjecture of quantum gravity \cite{Banks:1988yz,Giddings:1987cg,Lee:1988ge,Abbott:1989jw,Coleman:1989zu,Kamionkowski:1992mf,Holman:1992us,Kallosh:1995hi,Banks:2010zn,Krauss:1988zc}, which relies on Hawking radiation and appears to hold in string theory \cite{Banks:1988yz} and AdS/CFT \cite{Harlow:2018tng}, but may not be realized in asymptotic safety \cite{Eichhorn:2024rkc}, where phenomenological proposals of regular BH models have been made \cite{Bonanno:2000ep, Eichhorn:2022bgu, Platania:2023srt}.

A key direction for future work is to explore whether dynamical simulations of the theory \eqref{eq:action} are feasible \cite{Unluturk:2023qgk,Coates:2023dmz,Coates:2022qia,Silva:2021jya,Garcia-Saenz:2021uyv,Clough:2022ygm}, and, if so, to use them to investigate: (i) whether extremal regular BHs are the stable endpoint of collapse; (ii) whether they exhibit the Aretakis instability and its possible endpoint; and (iii) the stability of horizonless compact objects against the light-ring instability. A preliminary analysis presented in the \hyperref[suppmaterial]{Supplemental Material} indicates that black holes are linearly stable under radial perturbations. This analysis also shows that the vector field perturbations, despite implementing a non-linear constraint due to the lack of a kinetic term, are in fact tractable in a perturbation analysis. This provides a basis to study quasi-normal modes and stability under more general perturbations in the future.

Gravastars have long been conjectured to arise as solutions in semiclassical gravity \cite{Carballo-Rubio:2017tlh,Mazur:2001fv,Mottola:2022tcn,Mottola:2010gp}, particularly when trace anomaly effects are considered. Dimensional regularizations of the Gauss-Bonnet invariant are closely related to the effective action for the trace anomaly, at least in their scalar-tensor formulations \cite{Fernandes:2021dsb,Gabadadze:2023quw,Coriano:2022knl,Coriano:2022ftl}. The appearance of gravastar solutions in our theory warrants further investigation.

Finally, generalizing to spinning solutions is clearly an important direction and linked to the question whether extremality, and regularity, can be maintained in the course of dynamical evolution, which may lead to a change in spin.
Further, spinning up an already extremal BH may also generate a horizonless compact object, resulting in several phenomenologically interesting scenarios, e.g., a transition from two extremal BHs to a horizonless compact object through a merger with sufficiently large angular momentum, or a transition from a BH to a horizonless object when the spin increases through accretion. Both processes may lead to observational signatures in gravitational-wave interferometers and very-large-baseline interferometers for BH imaging.

\noindent \textbf{\textit{Acknowledgments.}}
We thank Christos Charmousis and Mokhtar Hassaine for discussions. This work is funded by the Deutsche Forschungsgemeinschaft (DFG, German Research Foundation) under Germany’s Excellence Strategy EXC 2181/1 - 390900948 (the Heidelberg STRUCTURES Excellence Cluster).
A.~E.~acknowledges the European Research Council's (ERC) support under the European Union’s Horizon 2020 research and innovation program Grant agreement No.~101170215 (ProbeQG).

\bibliography{references}

\onecolumngrid

\renewcommand{\theequation}{SM.\arabic{equation}}
\setcounter{equation}{0}
\section*{Supplemental Material}
\label{suppmaterial}
\subsection*{Field equations and stress-energy tensor}
The field equations of the theory given in Eq.~\eqref{eq:action} in the main text are
\begin{equation}
    G_{\mu \nu} = 8\pi T_{\mu \nu} = 8\pi (T_{\mu \nu}[A] + T_{\mu \nu}[B]),
\end{equation}
where the stress-energy tensor of each vector-field is defined as
\begin{equation}
    T_{\mu \nu}[A] = \frac{\ell^2}{8\pi}\mathcal{T}_{\mu \nu}[A], \qquad T_{\mu \nu}[B] = - \frac{\ell^2}{8\pi} \mathcal{T}_{\mu \nu}[B],\label{eq:Tmunugeneral}
\end{equation}
where
\begin{equation}
    \begin{aligned}
        \mathcal{T}_{\mu \nu}[W] = & -g_{\mu\nu} [-4 W^{\alpha} W^{\beta} R_{\alpha\beta}+8 W^{\alpha} W^{\beta} \nabla_{\beta}W_{\alpha}+2 \nabla_{\alpha}W^{\alpha} \nabla_{\beta}W^{\beta}+4 W^{\alpha} \nabla_{\beta}\nabla_{\alpha}W^{\beta}\\&-4 W^{\alpha} \nabla_{\beta}\nabla^{\beta}W_{\alpha}+2 \nabla_{\alpha}W_{\beta} \nabla^{\beta}W^{\alpha}-4 \nabla_{\beta}W_{\alpha} \nabla^{\beta}W^{\alpha}-3 W_{\alpha} W^{\alpha} W_{\beta} W^{\beta}+R W_{\alpha} W^{\alpha}]\\&-2 W^{\alpha} W_{\nu} R_{\mu\alpha}+2 W_{\alpha} W^{\alpha} R_{\mu\nu}-2 W^{\alpha} W_{\mu} R_{\nu\alpha}-4 W^{\alpha} W^{\beta} R_{\mu\alpha\nu\beta}-8 W_{\mu} W_{\nu} \nabla_{\alpha}W^{\alpha}\\&-2 W_{\nu} \nabla_{\alpha}\nabla^{\alpha}W_{\mu}+8 W^{\alpha} W_{\nu} \nabla_{\mu}W_{\alpha}+2 W_{\nu} \nabla_{\mu}\nabla_{\alpha}W^{\alpha}-2 W_{\mu} \nabla_{\alpha}\nabla^{\alpha}W_{\nu}\\&-4 \nabla^{\alpha}W_{\mu} \nabla_{\alpha}W_{\nu}+2 \nabla_{\mu}W_{\alpha} \nabla^{\alpha}W_{\nu}+2 \nabla_{\alpha}W^{\alpha} \nabla_{\mu}W_{\nu}+2 W^{\alpha} \nabla_{\mu}\nabla_{\alpha}W_{\nu}\\&+8 W^{\alpha} W_{\mu} \nabla_{\nu}W_{\alpha}+2 \nabla^{\alpha}W_{\mu} \nabla_{\nu}W_{\alpha}-4 \nabla_{\mu}W^{\alpha} \nabla_{\nu}W_{\alpha}+2 \nabla_{\alpha}W^{\alpha} \nabla_{\nu}W_{\mu}+2 W_{\mu} \nabla_{\nu}\nabla_{\alpha}W^{\alpha}\\&+2 W^{\alpha} \nabla_{\nu}\nabla_{\alpha}W_{\mu}-4 W^{\alpha} \nabla_{\nu}\nabla_{\mu}W_{\alpha}-12 W_{\alpha} W^{\alpha} W_{\mu} W_{\nu}+2 R W_{\mu} W_{\nu}.
    \end{aligned}
\end{equation}
The vector-field equations are given by
\begin{equation}
    \begin{aligned}
        &3 A_\alpha A^\alpha A^\mu + A_\alpha G^{\mu \alpha} + 2 A^\mu \nabla_\alpha A^\alpha - 2 A_\alpha \nabla^\mu A^\alpha = 0,\\&
        3 B_\alpha B^\alpha B^\mu + B_\alpha G^{\mu \alpha} + 2 B^\mu \nabla_\alpha B^\alpha - 2 B_\alpha \nabla^\mu B^\alpha = 0.
    \end{aligned}
    \label{eq:cov_VF_eqs}
\end{equation}
In addition, on-shell for our solutions, the vector fields are null, $A_\mu A^\mu = B_\mu B^\mu = 0$, so that it follows that $A_{\mu}A_{\nu}G^{\mu\nu}=0=B_{\mu}B_{\nu}G^{\mu\nu}$; and thus also $A_{\mu}A_{\nu}R^{\mu\nu}=0=B_{\mu}B_{\nu}R^{\mu\nu}$.

In standard Schwarzschild-type coordinates
\begin{equation}
    \dd s^2 = -f(r) \dd t^2 + \frac{\dd r^2}{f(r)} + r^2 \left( \dd \theta^2 + \sin^2\theta \dd \varphi^2 \right),
    \label{eq:SSS_usual_coordinates}
\end{equation}
for the solution given in Eq.~\eqref{eq:regular_bh}, we have
\begin{equation}
    \begin{aligned}
        &T^{t}_{\phantom{t}t}[A] = T^{r}_{\phantom{r}r}[A] = \frac{3\ell^2}{32\pi}\left[ \frac{q_a^2}{r^6}-\frac{64 q_a \ell^2 M^2}{\left(r^3 + 2 q_a \ell^2\right)^3}-\frac{8 M (q_a-2M)}{\left(r^3+2 q_a \ell^2\right)^2} \right],\\
        &T^{\theta}_{\phantom{\theta}\theta}[A] = T^{\varphi}_{\phantom{\varphi}\varphi}[A] = \frac{3\ell^2}{16\pi}\left[ -\frac{q_a^2}{r^6}+\frac{8 M \left((q_a-2M)r^6 + \ell^2 q_a (q_a+12M)r^3 - 2\ell^4 q_a^2 (q_a+2M)\right)}{\left(r^3 + 2 q_a \ell^2\right)^4} \right].
    \end{aligned}
\end{equation}
From the fact that $T_{\mu\nu}[A]$ and $T_{\mu\nu}[B]$ differ by an overall sign (cf.~Eq.~\eqref{eq:Tmunugeneral}), and that the terms proportional to $q_a$ in $A_{\mu}$ differ by an overall sign from those $B_{\mu}$, one might first expect that all terms proportional to odd powers of $q_a$ in $T_{\mu\nu}[A]$ agree in sign with the corresponding terms in $T_{\mu\nu}[B]$, and the terms  proportional to even powers of $q_a$ have the opposite sign. This expectation is born out for the leading-order, divergent term in the expansion in $r$, but not for the subsequent, finite terms. In fact,
\begin{equation}
    \begin{aligned}
        &T^{t}_{\phantom{t}t}[B] = T^{r}_{\phantom{r}r}[B] = \frac{3\ell^2}{32\pi}\left[ -\frac{q_a^2}{r^6}+\frac{64 q_a \ell^2 M^2}{\left(r^3 + 2 q_a \ell^2\right)^3}-\frac{8 M (q_a+2M)}{\left(r^3+2 q_a \ell^2\right)^2} \right],\\
        &T^{\theta}_{\phantom{\theta}\theta}[A] = T^{\varphi}_{\phantom{\varphi}\varphi}[A] = \frac{3\ell^2}{16\pi}\left[ \frac{q_a^2}{r^6}+\frac{8 M \left((q_a+2M)r^6 + \ell^2 q_a (q_a-12M)r^3 - 2\ell^4 q_a^2 (q_a-2M)\right)}{\left(r^3 + 2 q_a \ell^2\right)^4} \right].
    \end{aligned}
\end{equation}
For the total stress-energy tensor, divergent contributions cancel out and we get
\begin{equation}
    \begin{aligned}
        &T^{t}_{\phantom{t}t} = T^{r}_{\phantom{r}r} = - \frac{3\ell^2 q_a M}{2\pi\left(r^3+2 q_a \ell^2\right)^2},\\
        &T^{\theta}_{\phantom{\theta}\theta} = T^{\varphi}_{\phantom{\varphi}\varphi} = \frac{3 \ell^2 q_a M \left(r^3-q_a \ell^2\right)}{\pi  \left(r^3+2 q_a \ell^2\right)^3},
    \end{aligned}
\end{equation}
which agrees with the usual expression for the Hayward metric when $q_a = M$, see Ref.~\cite{Hayward:2005gi}.

\subsection*{Linear stability}
The black holes of the theory are linearly stable to radial perturbations. To prove this, we consider radial perturbations about the background metric in standard Schwarzschild-type coordinates
\begin{equation}
    \begin{aligned}
        &\dd s^2 = - \left(f(r) + \epsilon F_1(t,r) \right) \dd t^2 + \left(\frac{1}{f(r)} + \epsilon F_2(t,r) \right) \dd r^2 + r^2 \left( \dd \theta^2 + \sin^2\theta \dd \varphi^2 \right),\\&
        A_\mu \dd x^\mu = \left( a(r) + \epsilon \mathcal{A}_1 (t,r) \right) \dd t + \left( \frac{a(r)}{f(r)} + \epsilon \mathcal{A}_2 (t,r)  \right) \dd r,\\&
        B_\mu \dd x^\mu = \left( b(r) + \epsilon \mathcal{B}_1 (t,r) \right) \dd t + \left( \frac{b(r)}{f(r)} + \epsilon \mathcal{B}_2 (t,r)  \right) \dd r,
    \end{aligned}
    \label{eq:SSS_pert}
\end{equation}
where $f(r)$, $a(r)$ and $b(r)$ are the background quantities discussed in the main text, $F_{1,2}$, $\mathcal{A}_{1,2}$, and $\mathcal{B}_{1,2}$ are the perturbations of the metric and vector fields, and $\epsilon$ is a small bookkeeping parameter. We study the equations of motion to first-order in $\epsilon$, which are linear in the perturbations.

Following standard techniques in black hole perturbation theory, we choose to work with the perturbations in Fourier space
\begin{equation}
    F_{1,2} = \int \dd \omega e^{i \omega t} \tilde{F}_{1,2}(\omega,r), \quad \mathcal{A}_{1,2} = \int \dd \omega e^{i \omega t} \tilde{\mathcal{A}}_{1,2}(\omega,r), \quad \mathcal{B}_{1,2} = \int \dd \omega e^{i \omega t} \tilde{\mathcal{B}}_{1,2}(\omega,r),
\end{equation}
such that time derivatives are transformed into powers of $i \omega$.

We start by analysing the perturbation equations for $A_\mu$ given in Eq. \eqref{eq:cov_VF_eqs}. Both $t$ and $r$ components of the vector equations can be written in the form $x+i y = 0$, where $x$ and $y$ depend on the background quantities and perturbations. Thus, the real and imaginary parts of the equation should vanish separately. Imposing the vanishing of the imaginary part, we find
\begin{equation}
    \tilde{\mathcal{A}}_1 = \frac{(q_a-2r+2rf)\tilde{F}_1+f^2(q_a-2r-2rf)\tilde{F}_2}{8r^2f}, \quad \tilde{\mathcal{A}}_2 = \frac{f}{2r} \tilde{F}_2.
\end{equation}
Following exactly the same steps for the imaginary part of the $B_
\mu$ equations we find
\begin{equation}
    \tilde{\mathcal{B}}_1 = \frac{(q_b-2r+2rf)\tilde{F}_1+f^2(q_b-2r-2rf)\tilde{F}_2}{8r^2f}, \quad \tilde{\mathcal{B}}_2 = \frac{f}{2r} \tilde{F}_2.
\end{equation}

Upon using these expressions for the perturbations of $A_\mu$, and $B_\mu$, we find that the vector equations in Eq. \eqref{eq:cov_VF_eqs} reduce to two independent equations in total, which allow only the solution
\begin{equation}
    \tilde{F_1} = c_1 f, \qquad \tilde{F}_2 = 0, 
\end{equation}
where $c_1$ is an integration constant, which can be set to zero without loss of generality with a reparametrization of the temporal coordinate. Thus, we find that both metric perturbations vanish, and as a consequence, so do the vector perturbations. With all linear perturbations set to zero, this further implies there are no source terms for higher-order perturbations.

This result implies that the black holes are linearly stable, since all perturbations must vanish. We note that a very similar situation was observed in Ref. \cite{Fernandes:2021ysi}, also in the context of a Gauss-Bonnet theory, and is reminiscent of perturbations of this kind in GR, which must vanish due to Birkhoff's theorem.

\subsection*{Non-derivative curvature invariants}

As shown in \cite{Simpson:2023apa}, a necessary and sufficient condition for all non-derivative curvature invariants of a static, spherically symmetric metric to remain finite is that the Kretschmann scalar is everywhere bounded.
For the generalized metric given in Eq.~\eqref{eq:newregular}, the Kretschmann scalar is given by
\begin{equation}
    \begin{aligned}
        R_{\mu \alpha \nu \beta} R^{\mu \alpha \nu \beta} =& \frac{3M^2}{\eta ^6 \left(\eta +2 M^3 Q+r^3\right)^2} \bigg[ 27 \eta ^6+2 \eta ^5 \left(38 M^3 Q-13 r^3\right)-60 \eta ^3 \left(r^3-2 M^3 Q\right)^2 \left(2 M^3 Q+r^3\right)\\&+54 \eta  \left(r^3-2 M^3 Q\right)^4 \left(2 M^3 Q+r^3\right)+27 \left(r^3-2 M^3 Q\right)^4 \left(2 M^3 Q+r^3\right)^2\\&+3 \eta ^4 \left(92 M^6 Q^2-20 M^3 Q r^3+23 r^6\right)-27 \eta ^2 \left(r^3-2 M^3 Q\right)^2 \left(4 M^6 Q^2+12 M^3 Q r^3+r^6\right) \bigg],
    \end{aligned}
\end{equation}
where we have defined
\begin{equation}
    \eta = \sqrt{\left(2 M^3 Q+r^3\right)^2-8 \lambda  M^3 r^3}.
\end{equation}

The Kretschmann scalar might diverge in locations such that
\begin{equation}
    \eta + 2M^3 Q + r^3 = 0.
    \label{eq:condition_div_1}
\end{equation}
or
\begin{equation}
    \eta = 0,
    \label{eq:condition_div_2}
\end{equation}
For $Q>0$, which we assume in this work, Eq.~\eqref{eq:condition_div_1} has no solutions. On the other hand, Eq.~\eqref{eq:condition_div_2} has solutions when
\begin{equation}
    r^3 = 4 M^3 \left( \lambda \pm \sqrt{\lambda(\lambda-Q)} -Q/2\right),
\end{equation}
which are real and positive in $r$ only when $\lambda>0$ and $0<Q<\lambda$. Therefore, the spacetime in Eq.~\eqref{eq:newregular}, describes a regular object when $Q>\max\{\lambda,0\}$.
In the limit where $Q=\lambda$, the metric function in Eq.~\eqref{eq:newregular} reduces to a Heaviside function that joins a Schwarzschild metric and a de Sitter core, as given in Eq.~\eqref{eq:gravastar}.

\subsection*{Continuation of geodesics across $r=0$}
Ingoing radial null geodesics for the metric \eqref{eq:SSS_usual_coordinates} are described by
\begin{equation}
-f(r)\dot{t}^2 + f^{-1}(r)\dot{r}^2=0.
\end{equation}
Combined with the conservation of $e = - g_{tt} \dot{t} = f(r) \dot{t}$, we arrive at
\begin{equation}
\dot{t}= e f^{-1}(r)= e \left(1+ \frac{r^2}{Q\, M^2}+ \frac{r^4}{Q^2\, M^4} + \frac{(\lambda-Q)r^5}{2M^5 Q^3} \right)+ \mathcal{O}(r^6).
\end{equation}
for the metric function in Eq.~\eqref{eq:newregular}.
For $Q \neq \lambda$, the $r^5$-term in this expansion results in the geodesic not being a $C^{\infty}$ function at $r=0$, just as in the Hayward case discussed in \cite{Zhou:2022yio}. For $Q=\lambda$, when the spacetime transitions from an exact de Sitter core to a Schwarzschild exterior at the transition radius $r_{\rm trans}$, all uneven terms in the Taylor expansion of $\dot{t}$ are removed, and the geodesic is a $C^{\infty}$ function.

\end{document}